# A Mobile Management System for Reforming Subsidies Distribution in Developing Countries

Tarek El-Shishtawy
*Benha University, Faculty of Computers and Informatics, Egypt*
t.shishtawy@ictp.edu.eg

**Abstract**

This paper has a specific objective of being useful for showing how the advances in mobile technologies can help for solving social and political aspects involved in the reform of subsidies in developing countries. It describes the work done to build a mobile-based supportive network that integrates all subsidies partners: governmental, non-governmental organizations, merchants, and beneficiaries. One main contribution of this work is the setting of a framework for identifying the requirements of subsidies distribution information systems. In the proposed approach, seven domains were identified to build a Mobile Subsidizing Business Model (MSBM). Based on MSBM, detailed requirements were gathered in three stages, with each having its appropriate methodology. In this work, we focus on the layered architecture implementation of the subsidizing mobile system to breakdown the complexities, which are due to variations of mobile technologies, different business rules, and multiple distribution scenarios.

## 1. Introduction

Subsidy systems have been a mainstay of development countries government's long-term policy of promoting social equity and political stability. This policy attempts at sharing national natural resource wealth (Akhter, 2001). Food subsidy has been a major component of the social safety net for the poor, guaranteeing the availability of affordable staples, helping to reduce infant mortality and malnutrition. In spite of subsidize high cost, many specialists agree that the current 'manual' or "card" subsidies distribution systems are still not effective at reaching the poorest people. It reveals that the majority of benefits go to the non-needy. Another main problem in subsidizing efforts is the leakage. The larger the difference between the regulated price and the market price, the higher the incentive to leak. For example, in Egypt, measurement of subsidies is often difficult as they may be donated through many channels, including governmental, non-governmental and even political parties who offer non regular subsides. This represents additional political problem as these different channels can, in turn, affect the transparency of the subsidy and the political dynamics associated with revising or eliminating a subsidy (Simon, 2012).

Therefore, this work aimed to build a complete proof-of-concept mobile system to show that applying a gradual reform program is possible using technological means. The current system provides an effective, simple, easy-to-use, transparent, and efficient solution to deal with the problem of subsidies distribution. The proposed Subsidies Mobile System makes use of mobile phone as a simple, effective, and efficient tool for subsidiary distribution framework. To ensure wide acceptance of the system, many new added values features were implemented to encourage all key players to participate.

Subsidizing systems have multi-discipline requirements as they touch a wide range of people. Subsidizing by its nature has many socioeconomic factors that should be satisfied. Distribution of subsidizing rations or goods is a business that has to meet standard supply chain requirements. A complementary prospective is that mobile subsidizing system can be considered as a service targeting poor and less-educated people who have their own requirements. Also, offering services using mobile devices add more constraints that should be fulfilled. To build a successful multi-discipline system, we propose Mobile Business Subsidizing Model MBSB as a framework through which requirements can be gathered.

The remaining of this paper is organized into four sections: In the first section, we present the previous works of employing ICT in subsidize management, and the use of mobile technology in providing services in developing countries. This section also reviews indicators of mobile technology in Egypt. Second section introduces the proposed Mobile Subsidizing Business Model MSBM, and the data gathering methodologies. Results of analysis that details the components of MSBM are described in the third section. In the last section, we present the anatomy of the proposed layered-architecture design that fulfills the gathered requirements of the system.

2. **Previous works**

Subsidies distribution is a new discipline as a mobile application. Therefore, in this section we will review two related topics. The first is the more general problem of the use of ICT in controlling subsidizes distribution to enhance its effectiveness. Subsidizing activities can be considered services managed and offered through mobile devices. Therefore, the second topic gives examples of the mobile systems that supply services in developing countries. In the next section, the theoretical aspects of modeling mobile subsidizing business will be reviewed.

Multiple ICT systems have been developed to enhance the administration of subsidies distribution. The goal is to control corruption and leakage in the delivery mechanism of public service delivery systems. Corruption is related in functioning of governing bodies and system, Illiteracy, and lack of awareness. For example, Dhand et., al., (2008) built a Unified Ration Card database to computerize food grain supply chain. Ration card has a unique number and a barcode printed through that database. To increase the awareness, call center with a toll free number was adopted to take complaints from citizens and give any desired information. Another work (Pingale et., al., 2013) improved the transparency of food distribution system by building a website that remotely monitor the outlets and vehicles providing rations. Both systems do not add values from beneficiary's point of view. Prakash and Sivasankar (2012) do not ignore beneficiaries in their proposed system. They used web technology to build a direct communication between governing bodies and public. Nevertheless, many questions are raised concerning cost , suitability, and coverage of low income technology-illiterate citizens.

In many developing countries, food stamps are adopted for distributing food rations. Food stamps have changed radically with technology. Advanced systems use electronic transfers directly into accounts that can be drawn on for purchases (Josling, 2011). Beneficiary's credit card is charged and can be read by the store accepting the payment. The Smart Card Technologies are suitable for use in different programs including healthcare, banking, pension programs, micro-finance, insurance and transportation programs. For example, it was explored in Malawi in the broader context of social protection (Devereux and Vincent, 2010). It was concluded that there is great potential for the use of technology in delivering social protection, especially if employed at a national scale. The card is now used in several developing countries. For example, the card technology can be found in South Africa, Nigeria, Tanzania, Namibia, Vietnam, and Columbia, Card is affordable technological means because it allows needy citizens to access their basic needs through a quicker means like any other person in the society. The card also provides the government with means to obtain relevant data about distribution of basic needs to beneficiaries (Hulela, 2012). For smart cards to be useful, an infrastructure networks have to be adopted in distribution outlets. This may not be available in many developing countries especially in rural regions. Also, card technology does not offer awareness to beneficiaries about types of available services. Therefore, in our work we tested the mobile technology as a tool of providing subsidiaries for poor citizens. Mobile phone is a cheap communication tool even for poor citizens.

Mobile wireless infrastructure exists in almost every region makes mobile applications flexible enough to deploy anywhere. Communication offers immediate utility to any user, urban or rural, rich or poor. The keypad's familiarity, memory, extended battery life, and a compact size makes the mobile an affordable tool for rural environment. A report from the International Telecommunication Union (ITU, 2012) explains that the greatest impact of mobile communication is the increase of the number of people who are in reach of a telephone connection. In developing countries, mobile communication leads to increased exchange of information on trade and health services, and hence, it contributes in development goals. Many systems were concerned with mobile as a tool for supplying services in developing countries (Donner, 2008). Examples are health care, economics, e-learning, agriculture, and many other applications.

Most approaches to mobile use were directed towards the discipline of mobile-health services. In general, mobile phones have generally proven to be a beneficial means of improving the healthcare system (Upkar 2006 and WHO 2005). The applications include using Short Message Service (SMS) for diseases prevention, consultation services, medical data collection, and teaching medical students in rural communities. The midwife mobile phone project (Arul, 2010) was implemented in Indonesia to transmit health statistics to a central database. The system provides health advice and information to rural midwives. Another discipline focuses on economic development domain. Parikh and Lazowska (Parikh and Lazowska, 2006) designed a mobile application framework (CAM) to address the information needs of the rural developing world. Based on this framework, they presented a microfinance application that is based on capturing two dimensional barcodes on paper forms using the mobile phone camera. In other research (Javid and Parikh, 2006), the supply chain distribution was proposed to be enhanced by a mobile information system that can capture the details and locations of field transactions among rural sales routes. In the field of agriculture, Sirajul and Grönlund (Sirajul and Grönlund, 2007), introduced a gap analysis and stakeholder theory intended to provide timely and accurate market information to farmers in Bangladesh. In conclusion, the growing number of mobile users, low operating cost, and the widely-spread stable mobile communication infra-structure networks that covers most urban and rural regions are the motivations for selecting mobile phone as a solution.

## 3. Methodology

In order to develop successful systems, identifying user requirements is essential. These requirements have to be taken into consideration carefully as design issues by application developers. Electing requirements for the subsidize distribution system involves achieving many objectives such as technological, organizational and social objectives. In such enterprise systems, researchers (Kruchten, 2004 and Al-Debei et.al. 2010) emphasize the importance of modeling the business before eliciting its requirements.

The research methodology employed for collecting requirements includes two steps:
   A. In the first step, we built a business model for mobile subsidizing. The main business domains were identified based on existing business models. This was considered a blueprint of essential framework, which was enhanced through conducted surveys.
   B. In the second step detailed components of the model categories were discovered through interviews, literature reviews, and surveys,. This provided new insights that were not made explicit prior to the collection of the data.

The following two subsections describe the proposed business model for mobile subsidizing system, followed by methodologies for collecting requirements based on the proposed model.

### 3.1 Modeling Mobile Subsidizing Business

In the current work, we built a Mobile Subsidizing Business Model (MSBM).as a framework for collecting and structuring features of the proposed system. In general, Business Model (BM) helps to gain a comprehensive understanding of the application domain, its information problems, and the functional requirements that the system must satisfy. The Business Model can be defined as the ways in which an organization along with its stakeholders creates value for each party involved (Bouwman et. al., 2008). The presented subsidizing system overlaps with various business models.

Subsidizing activities can be considered services managed and offered through mobile devices, and therefore; the application should comply with Generic Mobile Service Business Model (Bouwman et al. 2008). More specifically, the proposed system collects information from beneficiary's database to determine his/her needs, and the system offers services based on user features. From this prospective, subsidiary services can belong to Context-Aware Mobile Service Model (Hegering et al., 2004, De Reuver and Haaker, 2009), in which the information collected about users are used to adapt automatically the behavior of services. Also, subsidiary requirements overlap with mobile commerce essential features: In m-commerce systems, application developers must consider various users to provide better services to attract them (Büyüközkan, 2009). General features of m-commerce systems such as simplicity, usability, flexibility, interface, speed, etc., should also be considered in building the mobile subsidizing system. Management of subsidiary logistics cannot be analyzed outside the context of Supply Chains Management SCM. SCM is defined as the integration of key business processes from end user through original suppliers that provides products, services, and information and hence adds value for customers and other stakeholders (Lambert et al., 1998). Management of subsidiary logistics includes integration of organizations and merchant distribution outlets so that subsidizes are distributed at the right quantities, to the right locations, and at the right time, in order to minimize system wide cost while satisfying service level requirements (Simchi et. al., 1999). Also, subsidizing activities hit many socioeconomic requirements which should be included for wide acceptance of the system.

In this work, we propose a Mobile Subsidizing Business Model MSBM as a frame work to capture the structure and dynamics of the target partners for which the system will be deployed. The model is sourced from existing business models, that are tuned and augmented by specific features related to the specific problem (e.g. socioeconomic requirements). The components of MSBM hit the following domains, shown in Figure 1:

1. Targeting
2. Socioeconomics
3. Supply Chain Business Logic
4. Organizational
5. Profitability
6. Functionality
7. Technological

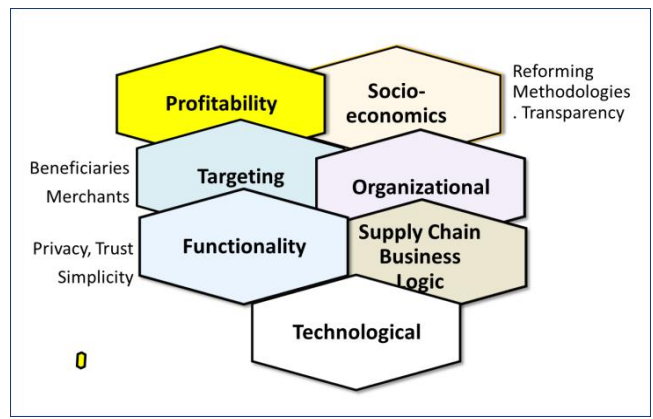

Figure 1: Components of Mobile Subsidizing Business Model

### 3.2 Requirements Gathering Methodologies

As our work is aimed at generating, rich findings, various methodologies were conducted to get the requirements' details based on MSBM. Detailed components were gathered in three stages:

**First Stage**: Socioeconomic factors were the essential driver of this work. We collected its detailed components through literature reviews in the area of reforming subsidizing efforts in Egypt. We benefited from the study previously done by National Centre for Social and Criminal Research (Nagwa, 2008) and the review study of Egyptian IDSC & WFP (2012) to understand and set the socioeconomic requirements.

**Second Stage**: We conducted series of semi-structured interviews with nearly 30 domain experts in 5 groups to cover the details of the proposed MSBM. To include a wide range of expertise, we selected different groups among various types of actors. The groups included policy makers in governmental organizations, NGO, mobile service operators, and practitioners who currently provide technological solutions (smart card) to cope partially with the problem. To guarantee a sufficient level of expertise in our sample groups, we selected senior level interviewees, i.e. CEOs, and consultancy. Table (1) shows interviewed groups, organizations, and their roles.

The objective of these interviews was to understand the main challenges that preclude the implementation of effective and efficient subsidy systems. A set of questions was prepared for each group based on the preliminary features of MSBM. During the interviews, experts were encouraged to prioritize features from their prospective and to suggest new requirements other than those shown on the preliminary list.

Table (1) : Interviewed Groups at CEO and Consultant levels

| Group | Organizations | Roles |
|---|---|---|
| Policy Makers Governmental Organizations | Ministry of supply and internal Trading | Responsible for achieving food security. Responsible for management of all subsidy governmental activities. |
| | Ministry of State for administration Development | Responsible for improving the management of state resources including subsidy new technological programmes. |
| NGO | RESALA Egyptian Bank of Foods | Two non-organizational organizations which have a long experience in providing clothes, foods medical and other services for poor families in Egypt. |
| Mobile Service Provider | Orange Research and Development Center-Egypt | One of the biggest mobile service providers in Egypt. |
| Practitioner | Private IT Practitioners | Two private Companies currently provide s smart card access as a partial solution to subsidy distribution in Egypt. |

**Third Stage**: In any application having the objective to satisfy its customer, one must focus on the needs of system users. We have conducted a survey study concerned with the problems and requirements from beneficiary and merchant prospective. As a result of the previous two stages, quota delivery scenarios (similar to those shown in Figures 3 and 4) were sketched and presented to interrogators to provide a near view of the system. A sample of 1280 users distributed over four regions was selected. The sample covered various ages as well as various education levels. The sample focused on two classes of user merchants (21%) and beneficiaries (79%).

Users were questioned to prioritize the importance of their needs. These needs relate to their views to the importance of general concepts like security, efficiency, simplicity, usability, speed, or Interface. In addition, respondents were asked to suggest their own expected added values when using the Mobile Subsidizing System.

### 4. Analysis Results based on MSBM

Based on the three-stages requirement gathering methodology described in the previous section, the detailed requirements of each MSBM domain were finally retained as detailed below,

#### *1- Targeting*

Targeting refers to the target groups of a subsidy services. We identified three target groups, shown in Figure 2:

A. Beneficiaries are the main target of the system. They are the Egyptian eligible citizens whom subsidy is intended to support. They are identified by their national IDs. For a beneficiary to join a system, he/she should have pin code, fixed address and a mobile number. Beneficiary may also represents a family with known number of associated persons.

B. Merchants have to be registered in the system in order to be able to distribute goods and services. Merchant is a concept that also includes all forms of service providers. Merchants can register as medical service providers, railway station offices, or merchants for food delivery. Geographical information is associated with merchants to analyze subsidiary distribution.

C. The third target group is the organization users. To aid privacy of the system, each organization has a number of users that administer all of the organization activities. Each user is assigned a role (e.g., administration, reporting, updating quota delivery, or managing beneficiaries).

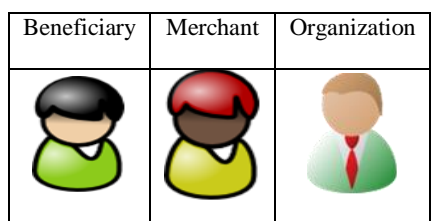

Figure 2: Actors of Target Groups

*2- Socioeconomic domain*

In socioeconomic domain, the following requirements were highlighted:

- Better targeting: efficient management is required to ensure reaching subsidies to needy people, and reducing subsidizes leakage.
- Transparency: It is an important social demand to unify different subsidiary channels, which lead to distribution transparency (Who gives aids to whom?)
- Reforming methodologies: One important requirement is to shift gradually towards cash compensation instead of offering goods with low prices. This eliminates having multiple prices for the same subsidizing item. The concept is applied through system functionalities to allow partial delivery of quotas and compensate non-delivered goods.

*3- Supply Chain Business Logic*

ICT helps to manage supply chain activities by offering information about what kind of product is demanded, and what is available in the warehouse. The system configured a simple supply chain network structure which involved tiers of suppliers (Governmental and Non-Governmental organizations), tiers of beneficiaries, and intermediate tiers of merchant outlets. Business processes were embedded in these tiers. The system implemented important business processes such as stock entry, stock transfer, quota definition, quota delivery, and other main functionalities of supply chain processes. The following describes important implemented processes

**Quota definition**: The system allows each governmental and nongovernmental organization to offer and distribute their subsidiary quotas such as food rations, butane containers for cooking needs, and fuel. Nongovernmental organizations can offer clothes, blankets, or even 'Adahi' meat quotas. The system allows quota to be personal or family based. Personal quota is computed for a single beneficiary- which is suitable for many subsidiary assistance forms such as gasoline, and non-governmental assistances. Family quota is computed for all family members represented by the beneficiary according to business rules defined by organizations. The family quota is the normal figure of food subsidiary delivery in Egypt.

An essential requirement is to allow organizations to determine dynamically the quota items. Items can change from one period to another. For each item, different costs are stored to compute the overall subsidizing cost and merchant profits. Participated organization sets quantities per beneficiary for each item, for example ½ Kgm. of oil, 1Kgm. of sugar, etc.).

**Distribution Schedule**: Another requirement of subsidize supply chain is to set the reputation time of quota distribution, i.e., yearly, monthly, weekly, daily, or once. The monthly-based quota is the normal subsidiary form in Egypt for food. However, the system allows other timely-based assistances.

**Quota Charging**: A main process of subsidizing supply chain is quota charging. Beneficiaries are notified when their balance is credited with new quotas. Since this feature is suitable for governmental quota, the system allows other organizations to set whether beneficiaries should be notified or not. The system charges all beneficiaries accounts with the specific quotas at right time according to the quota distribution schedule.

**Quota delivery**: Quota delivery is the central process of the proposed system that is completely managed by mobile devices of beneficiaries and merchants. Merchant can participate to deliver many types of quotas, as long as he/she is subscribed as a quota provider outlet. In some cases, beneficiaries don't need all items of the governmental rations. This may be due either its poor quality, or food habits in some regions. The proposed system introduced "partial delivery" or "refund" concept. It allows beneficiary to leave some items, and convert their subsidiary amount into cash transferred to his/her account.

The system supports many scenarios for quota distribution, based on the mobile device technology owned by beneficiary, and their technological knowledge level. It ranges from simple SMS to more sophisticated Java based mobiles. For example, Figure 3 shows SMS-Beneficiary Java-Merchant scenario, which assumes:

- Beneficiary requests all or part of his quota package
- Beneficiary owns non java enabled mobile and therefore communicate with SMS

- Beneficiary can read confirmation messages, and can reply with his pin code
- Merchant owns java enabled and has some IT skills to run a mobile application.

Another scenario is shown in Figure 4, where both of beneficiary and merchant own java enabled mobiles. The scenario assumes:

- Beneficiary requests all or part of his quota package
- Both of merchant and beneficiary own Java enabled mobile, and both have IT skills for running the mobile application.

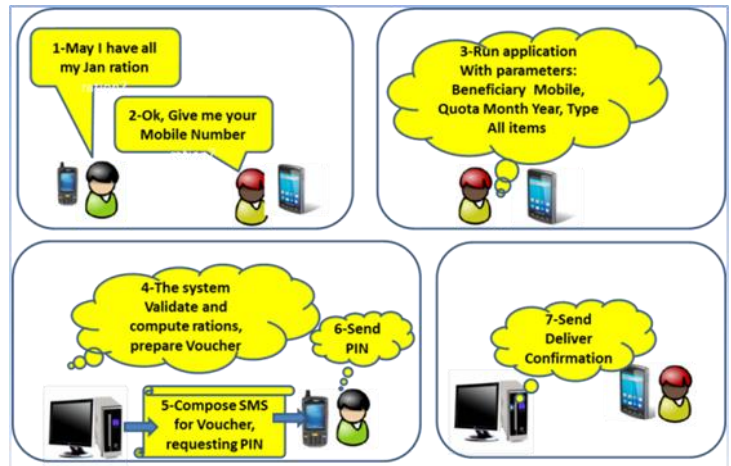

Figure 3: Quota Delivery using SMS-beneficiary, Java-Merchant Scenario.

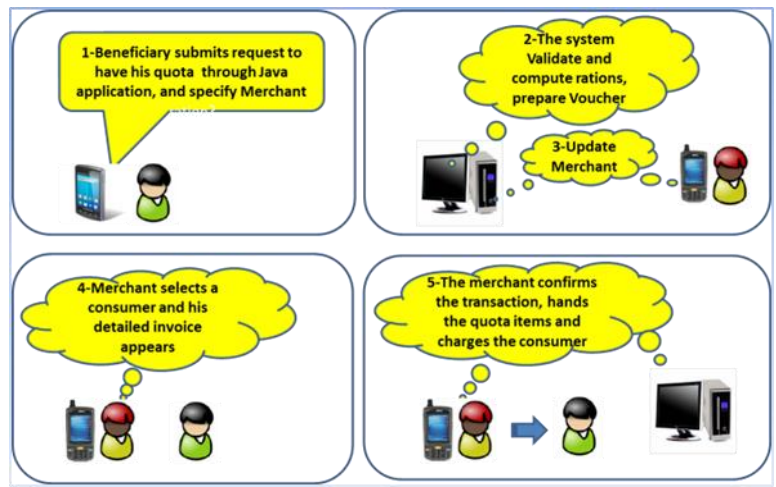

Figure 4: Quota Delivery using Java-beneficiary, Java-Merchant Scenario.

*4- Organizational Domain*

Interviews with governmental and NG organizations emphasized the following issues:

- Organizational openness: This indicates the ability of new organizations to join the system to offer subsidizing services to beneficiaries if they comply with the system rules. This implies designing a generic definition of services that target certain groups defined by organizations.
- Mobile service openness: This indicates the ability of the system to comply with different mobile service providers protocols.
- Governance: A main challenge is the governance. During interviews with organizational representatives, It was found that for this system to be realized, one important requirement is to assign a dominant role to governmental organization that has the access to the beneficiary database.

- Division of roles: one important requirement of NGO is to allow them to define their rules for targeting and to set their quota type, periodicity, and quantity.

*5- Profitability Domain*

One main goal of this work is to provide a public subsidize network that includes donating organizations, merchants, and beneficiaries. In this system, profitability is an essential dimension. All partners would like to know the profit that they will receive when they join the new system instead of relying on conventional way. In subsidizing business, financial profit is not the aim of all participants, however reducing overhead costs is a common objective. Then, in subsidizing systems profit is a concept that includes added values. To have a wide acceptance, the system is designed to provide a win-win situation for all partners. Through our analysis different groups assessed the importance of the following added values:

- From a community perspective, new concepts of socio-economic specialists for reforming subsidies strategies are implemented using mobile technology. The proposed system creates a cooperation environment for governmental and non-governmental organizations for distributing rations. To achieve the goal of having only one market price for a goods, the system implements a "refund" feature to move gradually towards cash substitution. Also, the proposed system supports statistical distribution reports for beneficiary habits over rural/urban regions and hence organizations can set alternative polices based on categories of required items.
- The added values for subsidizing organizations include the central facility of changing delivery rules, monitoring distribution process, and the ease of adding new subsidizing items. More subsidizing organizations can get benefit of the national beneficiary database and merchant distribution outlets.
- From beneficiary prospective, he/she is no longer tied to a specific merchant due to more flexible supply chain process, outlet, and offers better transparency of the subsidizing process. Also, using the same application, beneficiaries are subscribed automatically to other subsidizes offered by many organizations.
- Merchants have added values of better inventory control and monitoring with less paper work and shorter beneficiary queues. Using the same application, merchants can subscribe to distribute other subsidizes offered by many organizations. This will increase merchant profits.

*6- Functionality Domain*

During the third stage of requirements gathering, beneficiaries were asked to prioritize functionality requirements from their point of views. Among many functionality dimensions (flexibility, simplicity, usability, privacy, trust, speed, efficiency, and reliability), beneficiaries put privacy, simplicity, and trust as top priority requirements:

- Privacy: Most beneficiaries are not willing to give their subsidizers history to subsidizing organizations. For one reason or another, they considered this confidential information.
- Trust: Due to bad experience of beneficiaries with the current distribution system, they suggested new functionalities such as notification of quantities at the end of delivery processes, and ability to choose any merchant outlet as a supplier of quotas.
- Simplicity: As subsidizing systems target a wide range of consumers with various education levels, one important requirement is to keep the application as simple as possible. Many beneficiaries were worried about technological barrier. The current system requires only minimum knowledge of technologies. Simplicity is achieved through personalization. When beneficiary first joins, the system builds a user profile related to his/her mobile device platform, geographical, login information, and preferred merchant outlet. In most cases, requesting regular quotas requires only two clicks: one for requesting and the second for confirmation.

*7- Technology domain*

Technology domain includes issues that affect the quality of services offered by the system. Among analyzed features, accessibility got higher priority: One challenge of the mobile subsidizing system is the different platforms of mobile devices owned by users. Different device platforms have direct impact on both of communication protocols and user interfaces. Therefore, the architecture of the proposed system allows an open accessibility. In our implementation, a layered architecture of mobile server is proposed. The next section describes the role of presentation and communication layers for achieving required level of accessibility. System integration is a second required issue. It measures the extent to which a new service can be integrated into the existing technological infrastructure. In our implementation, business logic is separated from subsidizing basic delivery processes..

Reducing service cost, speed and security don't gain the expected importance through discussions with different groups. This is may be partially due to confidence of different groups that these requirements are essential in any information system.

Survivability which measures the ability of the system to operate even when the mobile set is out of coverage cannot be fulfilled in the current implementation. Implementing survivability requires specific mobile sets with minimum memory, which may be not available for large sectors of beneficiaries.

## 5. Layered Architectural Design

In this last section, we present a brief description of the layering architecture that implements the mobile subsidizing system. Layering is an architectural design pattern that structures applications so they can be decomposed into groups of subtasks. A layer represents a slice through the software architecture that represents a grouping of related functionalities. Layers restrict inter-system dependencies with the goal being to design a system that is more loosely coupled and thus easier to maintain. The software component within a given layer should access only components within its own layer or components in the layers beneath it. A common application of the layers design pattern organizes and defines the various layers within the problem domain based upon the responsibilities assigned to each layer. Layered architecture is chosen to allow generic design in which various user interfaces can be plugged into the system smoothly without change in the system core. Figure 5 shows the backend server side of the system. The system is partitioned into the following layers.

### 1) Data Access Layer

Data access layer consists of Database Servers. This bottom layer provides support for data access. This tier keeps data independent from application servers or business logic. The current system utilizes Oracle RDBMS within this environment to store the system concepts and their persistent information.

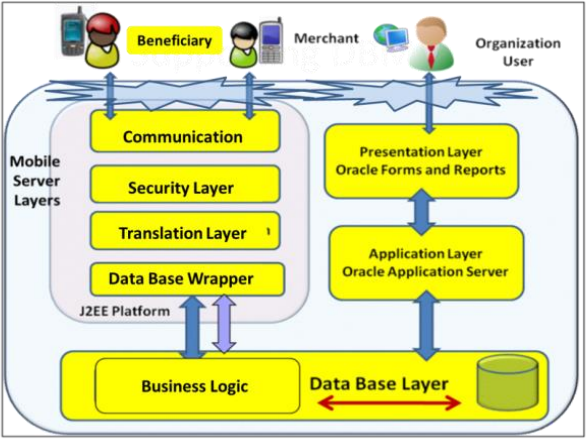

Figure 5: Layered Architecture of the mobile subsidizing system.

Figure 6 shows part of schema diagram of the delivery of quota items. The delivery process is stored as a beneficiary voucher. The voucher header carries the beneficiary id, quota id, and merchant id. Also, the header includes total quota prices for merchant and total quota price for beneficiary. This allows computation of the merchant profit, and helps for informing beneficiaries with their share price of quotas. The details of the voucher include the items and their rated (formal) quantities. Also, actual delivered quantities are stored to allow partial delivery of rations.

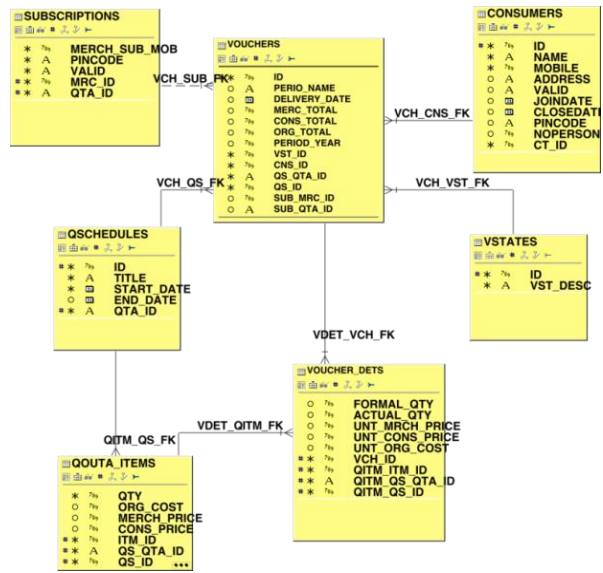

Figure 6: Part of Schema diagram showing beneficiary voucher

## *2) Business Logic Layer*

Implementation of business rules reside in database systems transparent to clients and their mobile devices. It controls application's functionalities by performing detailed processing subtasks. This middleware layer provides support for application specific business scenarios, as well as, data integrity rules. During the design phase, scenarios are converted into smaller sub-tasks, which may be requested by one or more scenarios. Table 2 shows part of implemented business logic sub-tasks. It is important to note that business logic layer implements sub-operations necessary to codify different scenarios. As will be shown, the translation layer responsibility is to decide when and how these sub-operations are called to fulfill requested scenario. For example, UpdateState can be issued after receiving a confirmation SMS from a beneficiary, or called when a button is clicked in Java-based mobile application.

Table 2: Examples of sub-tasks at business logic layer

| Business Subtask | Description |
| --- | --- |
| QueryQuota | Used to return available quota items for a beneficiary. |
| Charge_Quota | Issued to deliver quota items from specified merchant. It generates an invoice with state 'not delivered yet'. |
| UpdateQty | Issued to update a quantity of an item in invoice. |
| UpdateState | Issued to update the invoice state to be 'delivered' after delivery confirmation. |
| IsValidConsumer | Issued to validate consumer. |

Business logic is implemented with PL/SQL stored procedures, which are part of the database backend. Stored procedures allow developers to take advantage of procedural extensions to the standard declarative SQL syntax. It enables business logic to be stored as application logic in the database in compiled form. A stored procedure performs one or more specific task. This is similar to a procedure in other programming languages. A procedure has a header API and a body. The body consists or declaration section, execution section and exception section. For example, the header of the stored procedure which queries quota items is shown in Figure 7. It is used to return available quota items for a given consumer. The procedure accepts three input parameters identifying the quota, its schedule, and consumer (beneficiary). It returns a REF CURSORS type structure that contains Item_id, scheduled quantity, and consumer REF CURSORS allow record sets to be returned from stored procedures. The procedure also returns a Success flag as well as failure message describing its reasons.

```
QueryQuota (
        P_QTA_ID        IN      NUMBER ,
        P_QS_ID  IN     NUMBER,
        P_CNS_ID        IN      NUMBER,
```

|  | ITM_RECORDSET | OUT | TYPES.CURSOR_TYPE, |
|  | MYTXT | OUT | string, |
|  | MYSUCESS |  | OUT NUMBER ) |

Figure 7 APIs for Oracle Stored Procedures that query quota items

Figure 8 shows the implemented logic for charging quota items for a specific beneficiary.

```
PROCEDURE CHARGE_DETAIL (
        P_QTA_ID CHAR IN NUMBER
        P_QS_ID IN NUMBER
        P_CNS_ID IN NUMBER
        P_ID    IN NUMBER
        TXT OUT CHAR
        SUCESS   OUT NUMBER )  IS /* VOUCHER ID */

/* Local parameters */
LGROUP_BASED CHAR;
LPERSONS         NUMBER;
BEGIN
        SUCESS := 0;   /* ASSUME FAIL */
        TXT := ' cant charge quota items ';
/* Step 1 check that the qouta is personal or group based */
        SELECT GROUP_BASED INTO LGROUP_BASED
        FROM    QUOTAS
        WHERE   QUOTAS.ID = P_QTA_ID
/* STEP 2 fIND NUMBER OF PERSONS ASSIGNED TO QUOTAS */
        IF LGROUP_BASED = 'N' THEN
                LPERSONS := 1;
        ELSE
                BEGIN
                        SELECT NVL(NOPERSONS,1) INTO LPERSONS
                        FROM CONSUMERS
                        WHERE LPERSONS.ID = P_CNS_ID;
                EXCEPTION
                        WHEN OTHERS THEN
                                NOPERSONS := 1;
                END;
        END IF;
/* STEP 3 CHARGE THE VOUCHER DETAILS
INSERT INTO VOUCHER_DETS
(FORMAL_QTY, ACTUAL_QTY, UNT_MRCH_PRICE, UNT_CONS_PRICE,
UNT_ORG_COST, VCH_ID, QITM_ITM_ID, QITM_QS_QTA_ID, QITM_QS_ID)
        SELECT QTY*LPERSONS, QTY*LPERSONS, MERCH_PRICE, CONS_PRICE,
                ORG_COST, P_ID, ITM_ID, P_QTA_ID, P_QS_ID)
        FROM    QOUTA_ITEMS
        WHERE   QS_QTA_ID =  P_QTA_ID AND
                QS_ID = P_QS_ID;
        SUCESS := 1;   /* ASSUME FAIL */
        TXT := ' quota items CHARGED SUCCESS';
EXCEPTION
        WHEN OTHERS THEN NULL;
END;
```

Figure 8: An example of applying the business logic to charging beneficiary's quota

### 3) *Presentation Layers*

Presentation layer provides support for the interactions between the actors, or the users of the system, and the software system itself through the presentation of user interfaces. It is the topmost level of the application. It communicates with other tiers by outputting results to the browser/client and mobile application tiers. In the current design, two presentation layers do exist.

The first layer resides within forms server and called with normal web browser for subsidiary control at organization's side. Oracle Forms are used to build the organization's interfaces. Forms deliver a Rapid Application Development (RAD) environment and application deployment infrastructure to ensure that application automatically scale and perform over any network. Figure 9 shows an example of GUI for defining quota schedule and its components. User selects quota offered by his organization and set its periodicity. Second quotas schedule is set and user can put boundary of distribution terms of valid dates for each quota. Family quota can be limited to a maximum number of persons. Third, the items of each quota is defined, which can change from one period to another.

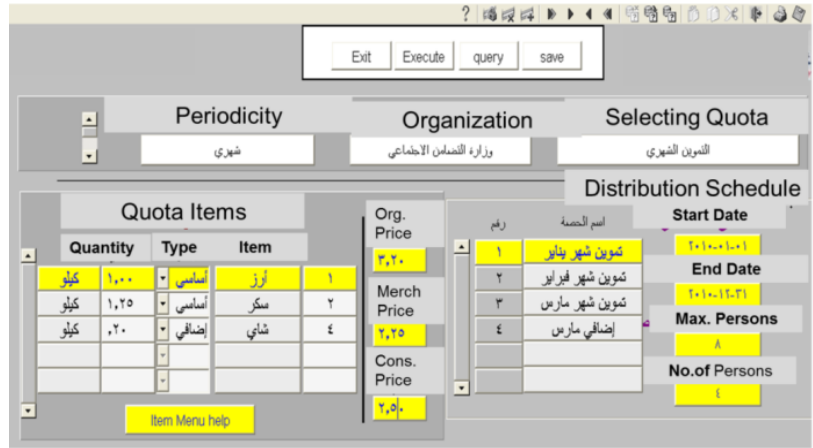

Figure 9: GUI for setting quota schedule by an organization user

The second presentation layer is implemented with java to run over merchant and consumer mobile phones. To minimize the negative impact of customization needed for different devices, a presentation layer is introduced. The layer carries specific platform features such as screen size and input type (touchscreen or keyboard).

*4) Mobile Server Layers*

Mobile server layers are responsible for implementing different quota distribution scenarios. As shown in figure 5, it is a middleware between data access layers and mobile phone sets. Other than communication and security layers, it includes database wrapper and translation layer.

*A. Database Wrapper*

A database wrapper object is suggested to isolate the mobile system from the details of typical organizational database and their business logic operations. Applications and applets can access databases via the JDBC API using pure Java JDBC technology-based drivers. With JDBC technology, businesses are not locked in any proprietary architecture, and can continue to use their installed databases and access information easily -- even if it is stored on different database management systems.

In the current design, database wrapper layer exploits a JDBC driver to establish a connection with the required database URL It creates an object instance of the connection, and manages different threads. On the mobile server side, stored procedures are called from JDBC programs. Database wrapper requests are executed by calling appropriate PL/SQL stored procedures. A call takes variable bind parameters as input parameters as well as output variables and creates an object instance of the Callable Statement class. The process is basically a remote procedure call (RPC) mechanism [Kshemkalyani, 2008]. The RPC mechanism conceptually works like a local procedure call, with the difference that the procedure code resides on a remote machine, and the RPC software sends a message across the network to invoke the remote procedure. It then awaits a reply, after which the procedure call completes from the perspective of the program that invoked it. For example, the statement

QueryQuota.registerOutParameter(4, OracleTypes. CURSOR);

Registers the fourth parameter passed to the stored procedure as an OUT parameter of type CURSOR_TYPE. Once the stored procedure has been executed, the values of the out parameters is obtained, parsed, assigned to appropriate objects for use in subsequent layer.

### B. Translation Layer

It was reported that the negative aspect of device variety is to raise development costs or to remove features which raise incompatibility (Holzer and Ondrus, 2011). In our work, we proposed a translation layer to burden technological changes of mobile sets owned by beneficiaries. Translation layer gained more importance since technologies are subject to evolution, and hence more different distribution scenarios will be feasible,

The translation layer is responsible for controlling the whole distribution processes according to beneficiary preference scenario. It keeps roadmaps of executing different scenarios in terms of callings to sub-business logic operations and communication protocols. According to the roadmap, the first functionality of this layer is to decide when and how a sub-logic operations defined at business logic layer (and buffered at database wrapper), is called to fulfill a specific scenario.

As shown in Figures 3 and 4, different scenarios require different communication protocols ranging from simple command line, SMS, to complicated java based protocols. Therefore the second functionality of the translation layer is to decide which communication protocol should be called to complete the distribution scenario.

The third important functionality of the translation layer is the preparation of data accepted from database wrapper in Java data structures to be communicated in suitable format for presentation layers reside at mobile phones. For example, when executing Java-based scenarios, relevant subsidizing information is uploaded to the beneficiary java-based phone when first joins. This minimizes communication cost in all subsequent transactions. On the other hand, non-java based phones require comprehensive messages to be composed by the translation layer.

## 6. Conclusion

This work demonstrates that mobile phones have generally proven to be a beneficial means of reforming the subsidies system. Using well established mobile networks, the system expands the subsidies coverage to include organizational, non-organizational entities, merchants, and beneficiaries. The geographical spread of merchant outlets, and centralized beneficiaries database allow many organizations to join smoothly to supply their subsidies. Mobile phones provided an efficient mean of communication to control subsidize leakage from consumer side. In conjunction with computer-based system at the organization side, the system added a second layer of leakage prevention control.

Although there is an increasing interest on ICT4D and its closely related activities of identifying user requirements, few researches concerned with gathering mobile subsidize requirements in the context of suitable Business Model. Therefore, this study represents an effort in this direction, by proposing a framework for the identification of user requirements. In the proposed approach, seven components were identified to build a Mobile Subsidizing Business Model (MSBM). The model was sourced from business related activities and includes Targeting, Socioeconomics, and Supply Chain Business Logic, Organizational, Profitability, Functionality, and Technological domains. Based on MSBM, detailed components were gathered in three stages. In the first stage, socioeconomic related articles were surveyed to sketch out the basic needs. Second, we conducted detailed semi-structured interviews with experts in four participating domains: Policy Makers, Governmental Organizations; NGO, Mobile Service Provider, and Practitioners. As a result of the previous two stages, scenario based descriptions were served as an initial view to end users. The third stage focused on end-user functionality requirements. Through a wide survey, beneficiaries and merchants were asked to priorities their functional needs.

The results show that the main issues for organizations are the openness, governance, and division of roles. Surprisingly, privacy and trust got higher priorities for beneficiaries, even before simplicity and added values. This may be due to the bad experience of beneficiaries with previous systems. This requires additional functionalities to build a trust between new systems and poor people. In our work, profitability is measured in terms of added values for each partner. In addition, lowering distribution costs and increasing the profit of merchants – when subscribed as service providers for NGO - were also considered.

Technically, many factors were considered through the design process. To ensure wide acceptance of the system, it is offered through many mobile platforms, ranging from java enabled devices to SMS based old phones. Different scenarios were implemented according to mobile technology. Through our work, we showed how different scenarios could be realized using layered architecture design, to allow different communication protocols as well as different mobile graphical interfaces.

The results of the presented system may be useful to practitioners in designing better accepted and satisfying systems, which take into account needs of various partners. In our work, we assert the importance of analyzing requirements in the context of the proposed MSBM, which is an effort towards better understanding of theoretical background of multi-domain subsidizing system. Finally, the implemented system can be useful for policy makers in developing countries. It shows that reforming many social and economic problems associated with subsidizing polices is possible when relying on modern ICT.

### 7. Acknowledgment

The author would like to acknowledge financial support of the project No. C2/S1/145 of EU-Egypt Innovation fund (EEIF), funded by the European Commission. The content of this paper is the sole responsibility of the author and can under no circumstances be regarded as reflecting the position of the European Union.